\documentclass[prl,reprint,superscriptaddress,longbibliography]{revtex4-1}
\usepackage{graphicx}
\usepackage{amssymb}
\usepackage{amsmath}
\usepackage[product-units = power]{siunitx}
\usepackage{amsfonts}
\usepackage{bbold}
\usepackage{bm}
\usepackage[version=3]{mhchem}
\usepackage[citecolor=blue,colorlinks=true]{hyperref}
\usepackage[dvipsnames]{xcolor}
\usepackage{nicefrac}
\usepackage{soul}

\DeclareSIUnit\BohrMagneton{$\mu_{\textrm{B}}$}
\DeclareSIUnit\formulaunit{f.u.}
\DeclareSIUnit\atomicunit{a.u.}
\DeclareSIUnit\arbunit{arb.unit}
\DeclareSIUnit\torr{Torr}
\DeclareSIUnit\counts{Counts}
\DeclareSIUnit\rlu{r.l.u.}

\definecolor{c1}{rgb}{0.2,0.8,0.3} % green
\definecolor{c2}{rgb}{0.8,0.3,0.2} % orangy
\definecolor{c3}{rgb}{0.85,0.33,0.1} % matlab red
\definecolor{c4}{rgb}{0.7,0.1,0.5} % pink-fuchsia
\definecolor{c5}{rgb}{0.3,0.2,0.8} % violet-blue

\definecolor{teal}{rgb}{0.0,0.5,0.5} % teal
\definecolor{navy}{rgb}{0.0,0.0,0.5} % navy
\definecolor{purple}{rgb}{0.5,0.0,0.5} % purple
\definecolor{darkorange}{rgb}{1.0,0.55,0.0} % dark orange
\definecolor{goldenrod}{rgb}{0.85,0.64,0.125} % golden rod
\definecolor{celticsgold}{rgb}{0.54,0.43,0.3} % celtics gold
\definecolor{hotpink}{rgb}{1.0,0.41,0.7} % hot pink
\definecolor{mediumvioletred}{rgb}{0.78,0.08,0.52} % medium violet red
\definecolor{turquoise}{rgb}{0.25,0.875,0.81} % turquoise
\definecolor{darkturquoise}{rgb}{0.0,0.8,0.82} % dark turquoise
\definecolor{forestgreen}{rgb}{0.13,0.54,0.13} % forest green
\definecolor{seagreen}{rgb}{0.19,0.54,0.34} % sea green
\definecolor{celticsgreen}{rgb}{0,0.48,0.2} % celtics green
\definecolor{steelblue}{rgb}{0.27,0.51,0.70} % steel blue
\definecolor{dodgerblue}{rgb}{0.12,0.56,1} % dodger blue
\definecolor{mediumblue}{rgb}{0.0,0.0,0.8} % medium blue
\definecolor{warriorsblue}{rgb}{0,0.412,0.71} % warriors blue
\definecolor{royalblue}{rgb}{0,0.324,0.73} % royal blue
\definecolor{hornetspurple}{rgb}{0.11,0.066,0.375} % hornets purple
\definecolor{yellowgreen}{rgb}{0.68,1.0,0.18} % yellow green
\definecolor{mediumseagreen}{rgb}{0.23,0.70,0.44} % medium sea green
\definecolor{wheat}{rgb}{0.96,0.87,0.70} % wheat

\usepackage{hyperref}
\hypersetup{
    bookmarks=true,         % show bookmarks bar?
    bookmarksopen=true,
    unicode=true,          % non-Latin characters in Acrobatss bookmarks
    pdftoolbar=true,        % show Acrobatss toolbar?
    pdfmenubar=true,        % show Acrobatss menu?
    pdffitwindow=false,     % window fit to page when opened
    pdfstartview={FitH},    % fits the page bounding box to the window
    pdfborder={0 0 0},      % width of PDF link border
    pdftitle={Multiple-magnon excitations shape the spin spectrum of cuprate parent compounds},    % title
    pdfauthor={Davide Betto et al.},     % author
    pdfsubject={Multiple magnons in layered cuprates},   % subject of the document
    pdfcreator={Davide Betto},   % creator of the document
    pdfkeywords={RIXS, Spin Waves, Magnons, Cuprates}, % list of keywords
    pdfnewwindow=true,      % links in new PDF window
    colorlinks=true,       % false: boxed links; true: colored links
    linkcolor=red,          % color of internal links (change box color with linkbordercolor)
    citecolor=mediumseagreen,        % color of links to bibliography
    filecolor=magenta,      % color of file links
    urlcolor=steelblue,           % color of external links
    linktocpage=true,       % makes the page numbers instead of the text to be link in the Table of contents
    hyperindex=true,        % add links to the table of contents
    pdfpagelabels=true,
    plainpages=false,
}

\begin{document}

%Title of paper
\title{Multiple-magnon excitations shape the spin spectrum of cuprate parent compounds}

\author{Davide Betto}
\email[Email address: ]{davide.betto@esrf.fr}
\affiliation{European Synchrotron Radiation Facility, B.P. 220, 38043 Grenoble, France}
\author{Roberto Fumagalli}
\affiliation{Dipartimento di Fisica, Politecnico di Milano, piazza Leonardo da Vinci 32, 20133 Milano, Italy}
\author{Leonardo Martinelli}
\affiliation{Dipartimento di Fisica, Politecnico di Milano, piazza Leonardo da Vinci 32, 20133 Milano, Italy}
\author{Matteo Rossi}
\affiliation{Dipartimento di Fisica, Politecnico di Milano, piazza Leonardo da Vinci 32, 20133 Milano, Italy}
\author{Riccardo Piombo}
\affiliation{ISC-CNR and  Dipartimento di Fisica, Universit\`{a} di Roma ``La Sapienza'', p.le Aldo Moro 5, 00185 Roma, Italy}
\author{Kazuyoshi Yoshimi}
\affiliation{The Institute for Solid State Physics, The University of Tokyo, Kashiwa-shi, Chiba, 277-8581, Japan}
\author{Daniele Di Castro}
\affiliation{Dipartimento di Ingegneria Civile e Ingegneria Informatica, Universit\`{a} di Roma Tor Vergata,, Via del Politecnico 1, 00133 Roma, Italy}
\affiliation{CNR-SPIN, Universit\`{a} di Roma Tor Vergata, Via del Politecnico 1, 00133 Roma, Italy}
\author{Emiliano Di Gennaro}
\affiliation{Dipartimento di Fisica ``E. Pancini'', Universit\`{a} degli Studi di Napoli ``Federico II'', Complesso Monte Sant'Angelo via Cinthia, 80126 Napoli, Italy}
\affiliation{CNR-SPIN, Complesso Monte Sant'Angelo via Cinthia, 80126 Napoli, Italy}
\author{Alessia Sambri}
\affiliation{CNR-SPIN, Complesso Monte Sant'Angelo via Cinthia, 80126 Napoli, Italy}
\author{Doug Bonn}
\affiliation{Department of Physics and Astronomy, University of British Columbia,Vancouver, British Columbia V6T 1Z1, Canada}
\author{George A. Sawatzky}
\affiliation{Department of Physics and Astronomy, University of British Columbia,Vancouver, British Columbia V6T 1Z1, Canada}
\author{Lucio Braicovich}
\affiliation{European Synchrotron Radiation Facility, B.P. 220, 38043 Grenoble, France}
\author{Nicholas B. Brookes}
\affiliation{European Synchrotron Radiation Facility, B.P. 220, 38043 Grenoble, France}
\author{Jos\'{e} Lorenzana}
\email[Email address: ]{jose.lorenzana@uniroma1.it}
\affiliation{ISC-CNR and  Dipartimento di Fisica, Universit\`{a} di Roma ``La Sapienza'', p.le Aldo Moro 5, 00185 Roma, Italy} 
\author{Giacomo Ghiringhelli}
\email[Email address: ]{giacomo.ghiringhelli@polimi.it}
\affiliation{Dipartimento di Fisica, Politecnico di Milano, piazza Leonardo da Vinci 32, 20133 Milano, Italy} 
\affiliation{CNR-SPIN, Dipartimento di Fisica, Politecnico di Milano, piazza Leonardo da Vinci 32, 20133 Milano, Italy}

\begin{abstract}
 Thanks to high resolution and polarization analysis, resonant inelastic x-ray scattering (RIXS) magnetic spectra of \ce{La2CuO4}, \ce{Sr2CuO2Cl2} and \ce{CaCuO2} reveal a rich set of properties of the spin $\nicefrac{1}{2}$ antiferromagnetic square lattice of cuprates. The leading single-magnon peak energy dispersion is in excellent agreement with the corresponding inelastic neutron scattering measurements. However, the RIXS data unveil an asymmetric lineshape possibly due to odd higher order terms. Moreover, a sharp bimagnon feature emerges from the continuum at ($\nicefrac{1}{2}$,0), coincident in energy with the bimagnon peak detected in optical spectroscopy. These findings show that the inherently complex spin  spectra of cuprates, an exquisite manifestation of quantum magnetism, can be effectively explored by exploiting the richness of RIXS cross sections. 
 \end{abstract}
\pacs{not.a.number}
\date{\today}

\maketitle

%\section{Introduction}
\label{sec:introduction} The spin $\nicefrac{1}{2}$ antiferromagnetic two-dimensional (2D) square lattice is one of the best studied quantum systems and represents a benchmark for quantum magnetism. Notably, it depicts the spin ground state arrangement in the \ce{CuO2} planes, common to all high-temperature superconducting layered cuprates, when no doping charge is present  and the antiferromagnetic order impedes charge transport and energetic magnons dominate the spin spectra \cite{Coldea2001,Headings2010}. Upon doping, long range antiferromagnetism is substituted by superconductivity but short-range in-plane spin correlations survive, giving rise to damped magnons of comparably high energy \cite{Braicovich2010,Dean2013,Minola2015,Peng2018}. Indeed, spin fluctuations are considered to be a main ingredient of the Cooper pairing ``glue'' in these materials \cite{Scalapino2012}, as suggested by the correlation between $T_\textrm{c}$ and the exchange interaction $J$ in certain cuprate families \cite{Moreira2001,Ofer2006,Peng2017,Ivashko2019,Grzelak2020}.
The spectrum of magnetic excitations has been extensively used to experimentally determine the coupling parameters with physical importance  \cite{Singh1989a,LorenzanaSawatzky1995,Lorenzana1995,Lorenzana1999,Coldea2001,Pavarini2001,Lorenzana2005,Headings2010,DallaPiazza2012,Peng2017,Wang2018}, such as $J$, the hopping integral $t$ and the Coulomb repulsion $U$.

The magnon dispersion is traditionally measured by inelastic neutron scattering (INS) and reproduced, within the linear spin-wave theory, by an improved Heisenberg model that includes higher-order terms  \cite{Headings2010,Plumb2014,Peng2017}. In the last decade resonant inelastic x-ray scattering (RIXS)  \cite{Ament2011} has proven to be a valid alternative to INS, in particular for cuprates  \cite{Braicovich2010,LeTacon2011,Dean2012,Minola2015,Peng2018} and other transition metal compounds with large superexchange coupling  \cite{Kim2012,Betto2017,Fabbris2017,Lu2018,Gretarsson2019}. 

Neutrons interact only with the electrons' spin, not with the charge, and with the atomic nuclei, making the theoretical treatment of the scattering cross-sections rather straightforward  \cite{Squires2012}. Consequently, once the phononic background is duly subtracted, the interpretation of the INS experimental spectra in terms of magnetic scattering function is in principle simple but, at the same time, it misses part of the richness of the many-body problem. Instead RIXS allows for a wide energy loss range measured at constant resolving power. Compared with INS it can profit from a much larger cross sections and incident flux but requires a more involved theoretical analysis  \cite{Jia2016}\footnote{See the Supplementary Material \cite{SupplementaryMaterial} for a pedagogical discussion of magnetic scattering in RIXS and neutrons
and the distinction between spin-conserving ($\Delta S=0$) and non spin-conserving
($\Delta S=1$) scattering processes.}, with less stringent selection rules. Therefore, RIXS has the potential to provide more information on the problem if one is able to disentangle the complexity of the spectra by exploiting good energy resolution and the analysis of the scattered x-rays' polarization (polarimetric analysis)  \cite{Braicovich2014,Fumagalli2019}.

\begin{figure*}[tb]
\centering
\includegraphics[width=\textwidth]{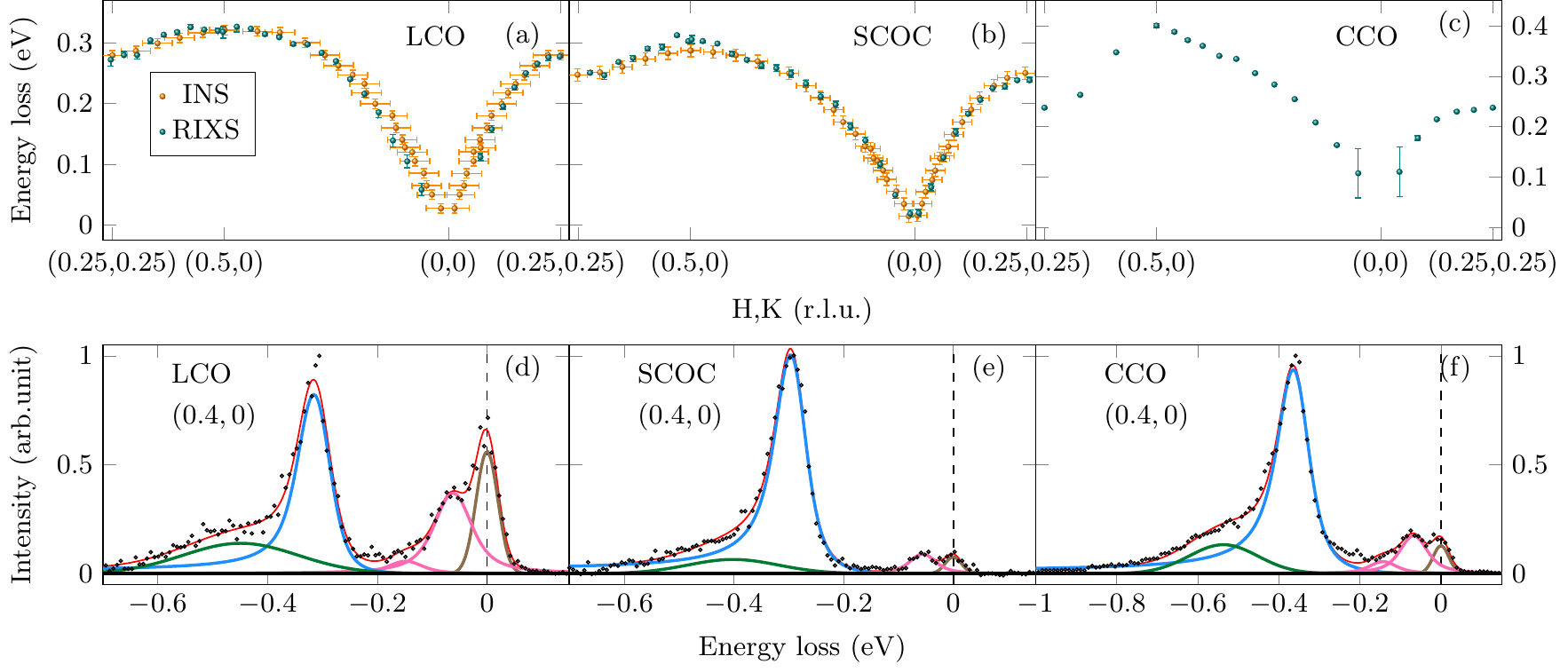}
\caption{(a-c): Single magnon dispersion determined by
RIXS for the three compounds and by INS for SCOC and
LCO (Refs.\,\onlinecite{Headings2010} and \onlinecite{Plumb2014}, respectively).
(d-f): RIXS spectra at (0.4,0) measured with  $\pi$ incident photon polarization (black circles) and
their main constituents obtained by a phenomenological fitting (red line): Gaussian elastic peak (brown), two resolution limited phonon contributions (pink), single magnon with Fano lineshape (blue), even order multimagnons (green).
}
\label{fig:dispersions}
\end{figure*}

These differences in the cross sections stimulated a close comparison of the two techniques, and some doubts were raised on the possibility of deriving the actual spin dynamical structure factor $S(\mathbf{q},\omega)$ from RIXS data. Indeed, in the very first RIXS work Braicovich \emph{et al.} \cite{Braicovich2009} had already shown that in \ce{La2CuO4} the single magnon energy dispersion in INS and RIXS coincide almost perfectly, a fact that has been recently confirmed more extensively \cite{Robarts2020}. However, Plumb \emph{et al.} \cite{Plumb2014} pointed out some discrepancies in the magnetic excitation spectrum of \ce{Sr2CuO2Cl2} close to ${\bf q}=(\nicefrac{1}{2}, 0)$ (X-point) of the two-dimensional (2D) Brillouin zone  \footnote{Throughout this
work we disregard the out-of-plane momentum and use reciprocal lattice units (\si{\rlu}) ($2\pi/a$, $2\pi/b$) where $a=b$ are the in-plane lattice parameters of the \ce{CuO2} planes.}, where the RIXS-derived magnon energy exceeds that of INS by $\sim
\SI{25}{\milli\electronvolt}$, i.e., about \SI{10}{\percent}. In this Letter, we start from this single magnon issue and unveil a richer scenario for the RIXS data. We report high-resolution RIXS measurements on a \ce{La2CuO4} (LCO) thin film ($\sim \SI{100}{\nano\metre}$-thick),  \ce{Sr2CuO2Cl2} (SCOC) crystals and a \ce{CaCuO2} (CCO) thin film,
with special emphasis on the low-energy (magnetic) portion of the spectra and on the comparison with the most recent INS data on the same compounds, where available.  A polarimetric analysis at selected momentum points, supplemented with theoretical computations, allows us to constrain the symmetry of the different contributions to the RIXS lineshape.

%\section{Experimental details}
\label{sec:experimental_details}

The RIXS spectra have been acquired using the \mbox{ERIXS} spectrometer of the ID32
beamline \cite{Brookes2018} at the European Synchrotron ESRF, which includes the polarimeter used
for the analysis of the polarization of the scattered
light  \cite{Braicovich2014,Fumagalli2019}. The x-ray energy was tuned to the Cu $L_3$
edge, at about \SI{931}{\electronvolt}. The incoming x-rays were polarized either
parallel ($\pi$) or perpendicular ($\sigma$) to the scattering plane (see SM Fig. S1 \cite{SupplementaryMaterial}). The total energy resolution was $\sim \SI{47}{\milli\electronvolt}$ for
the LCO and CCO, $\sim \SI{32}{\milli\electronvolt}$ for the SCOC and $\sim \SI{65}{\milli\electronvolt}$ for the polarimetric spectra. We mapped the magnon dispersion for the three compounds with $\pi$
polarization along the $(\nicefrac{1}{4}, \nicefrac{1}{4})$ $\rightarrow$
$(\nicefrac{1}{2}, 0)$ $\rightarrow$ $(0, 0)$ $\rightarrow$ $(\nicefrac{1}{4},
\nicefrac{1}{4})$ path in reciprocal space. %For further details on the measurements and the samples, see the Methods section in the SM \cite{SupplementaryMaterial}.

\label{sec:results}

Fig.\,\ref{fig:dispersions} shows the low energy-loss portion of selected spectra (see SM Fig.\,S2 for the complete set of data \cite{SupplementaryMaterial}). Each spectrum has been decomposed by phenomenological multi-peak fitting into an elastic line at zero energy loss, a phonon contribution and its overtone, a Fano lineshape (comprising a leading single magnon peak and a multimagnon tail) and an additional multimagnon peak, in order of increasing energy loss, as shown in panels (d-f) of Fig.\,\ref{fig:dispersions}. The elastic and the multimagnon peaks were modeled using Gaussian functions, while for the phonon peaks we employed a Lorentzian shape convoluted with the experimental resolution. For the additional multimagnon contribution, the choice of lineshape is not crucial since the spectrum is very broad energy-wise. For the single magnon peak and its tail, we found that the Fano asymmetric function gives the best results in the fitting procedure and its implications will be discussed below. We emphasize that the peak is not resolution limited for $q>\SI{0.35}{\rlu}$ along the [1,0] direction and that a Fano lineshape gives much better results across the whole Brillouin zone, especially on the low-energy side of the peak, as compared to a more conventional damped oscillator function \cite{Lamsal2016,Peng2018}.
The spectra close to ${\bf q}=(0, 0)$ are not shown because  the elastic component is too intense there and hinders the determination of the loss features.
All three compounds show the same asymmetric lineshape of the main peak, indicating that this is a common feature of 2D spin-$\nicefrac{1}{2}$ lattices and of cuprates.

The extracted single magnon peak positions are shown in Fig.\,\ref{fig:dispersions}\,(a-c) and compared to the INS results taken from literature when available
\cite{Headings2010,Plumb2014}. The LCO dispersion is almost superimposed to the INS data over the whole Brillouin zone  in agreement with previous literature \cite{Braicovich2010,Dean2012,Robarts2020}. In SCOC, the agreement is very good everywhere except close to the X-point. However, the better resolution of our RIXS data with respect to those of Ref.\,\onlinecite{Guarise2010}, previously used in Ref.\,\onlinecite{Plumb2014} for the comparison, allows us to better model the lineshape and to reduced the energy difference to $\sim \SI{15}{\milli\electronvolt}$. %depending on details of the lineshape modeling
The small difference at $(\nicefrac{1}{2}, 0)$ is mainly due to the inadequacy of a single peak to reproduce the actual spectral shape, whose determination is more challenging for INS than RIXS in this region of the momentum space where the scattering intensity is particularly low. Indeed, the X-point is characterized by a series of very interesting anomalies (weakening and broadening of the leading magnon peak, emergence of high energy tail) both in theoretical studies\cite{Sandvik2001,Sandvik2017},
and in physical realizations of spin-$\nicefrac{1}{2}$ square-lattice antiferromagnetic systems 
irrespective of the actual exchange energy scale  \cite{Christensen2007,Tsyrulin2009,Headings2010}. We can thus conclude that no significant discrepancy between INS and RIXS single magnon dispersion remains if the data are measured with adequate energy resolution and statistical quality and are analysed with the proper lineshape.

\begin{figure}
\centering
\includegraphics[width=\columnwidth]{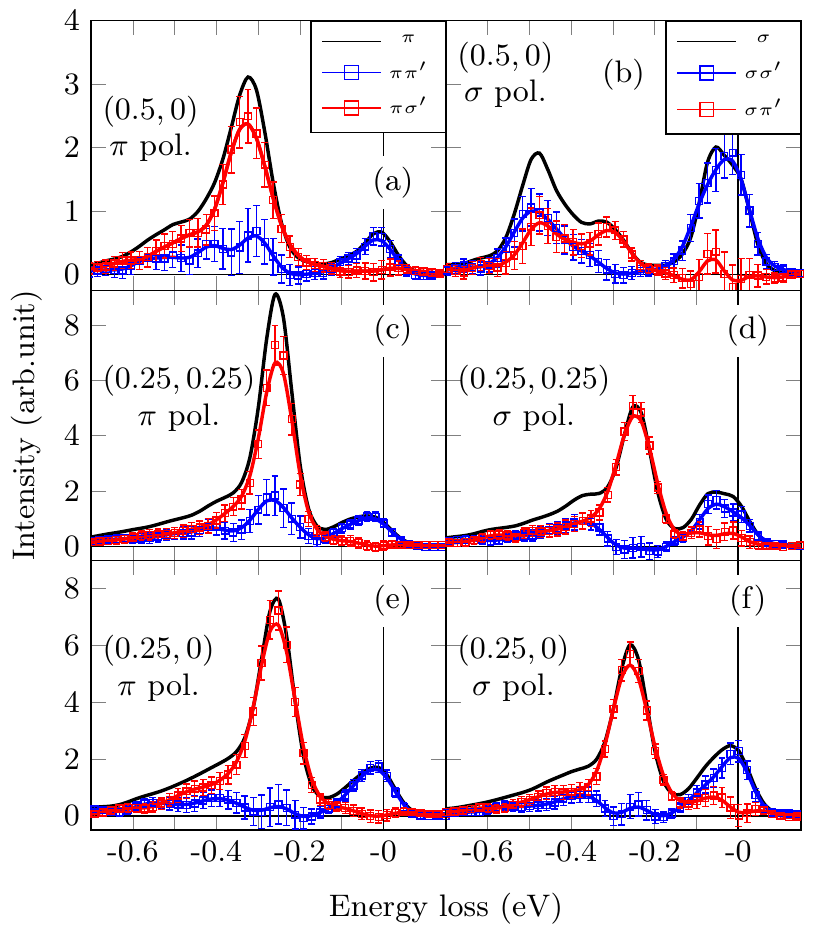}
\caption{Polarimetric spectra for SCOC at different ${\bf q}$. The red and blue solid lines are the
result of a 3-points adjacent-averaging
 of the experimental data points (squares). $\pi'$ and $\sigma'$ indicate the scattered polarizations.
%The vertical dotted line indicates the maximum of the
%one-magnon peak in neutrons  \cite{Plumb2014}.
 %The IR data on the first plot (from Ref.\,\onlinecite{Perkins1993}) show the IR absorption spectrum for SCOC.
 } \label{fig:polarimeter}
\end{figure}

The richness of the RIXS spectra invites to go beyond the traditional analysis made on INS data and to better exploit the complexity of the RIXS cross sections. In that spirit, we acquired RIXS spectra of SCOC with
analysis of the scattered light polarization at three different ${\bf q}$ values (Fig.\,\ref{fig:polarimeter}). Measurement methods, analysis and spectral assignments were made as in  Refs.\,\onlinecite{Braicovich2014,Fumagalli2019}. Figure~\ref{fig:polarimeter}(a) confirms that the main peak used to draw the single magnon dispersion of Fig.\,\ref{fig:dispersions} has crossed polarization character $\pi\sigma'$  (the prime indicates the scattered x-rays polarization). The rotation of the photon polarization after the scattering process, which implies a transfer of angular momentum, is needed for an odd number of magnons to be excited, i.e., for $\Delta S = 1$ spin flip process. Conversely the parallel polarization scattering channels must correspond to $\Delta S = 0$ spin conserving excitations, i.e., to an even number of magnons simultaneously excited \footnote{Two-magnon excitations can also contribute in crossed polarization
as in Raman but our exact diagonalization calculations show that their energies lie at higher energies.}.
With this in mind, the polarimetric data appear immediately of non-trivial complexity: they disprove the simplistic assignment of the high energy tail to two-magnons only and they reveal that parallel polarization spectra are different when $\pi$ or $\sigma$ incident polarization is used. The latter is particularly evident by comparing the blue curves of Fig.\,\ref{fig:polarimeter} in panels of the same row, and is very striking at the X-point.
It is often assumed that, in $\sigma\pi'$ or $\pi\sigma'$ polarization ($\Delta S=1$ excitations), the scattering involves a single on-site spin-flip operator
$\hat{S}^\pm_{\bf r}$, leading to a RIXS spectrum  proportional to the transverse magnetic structure factor $S^{\perp}({\bf q},\omega)$, {\em irrespective of the relative orientation between electric field and lattice}. From theoretical studies the 
latter is known to consist of a single magnon peak and a continuum of odd number of magnons developing above it \cite{Sandvik2001,Igarashi2005,Lorenzana2005,Igarashi2012}, with the latter having maximum relative spectral weight (\SI{40}{\percent}) at (\nicefrac{1}{2},0)  \cite{Sandvik2001,Sandvik2017} and around \SI{21}{\percent} weight on average in the
whole Brillouin zone \cite{Lorenzana2005}. The crossed polarization lineshapes at $(\nicefrac{1}{4}, \nicefrac{1}{4})$ and $(\nicefrac{1}{4}, 0)$ [red in Fig.~\ref{fig:polarimeter}(c,d,e,f) are consistent with these predictions. Furthermore, the $\pi\sigma'$ polarization at $(\nicefrac{1}{2}, 0)$  shows relatively more weight in the continuum as expected from the theory.

\begin{figure}
\centering
\includegraphics[width=\columnwidth]{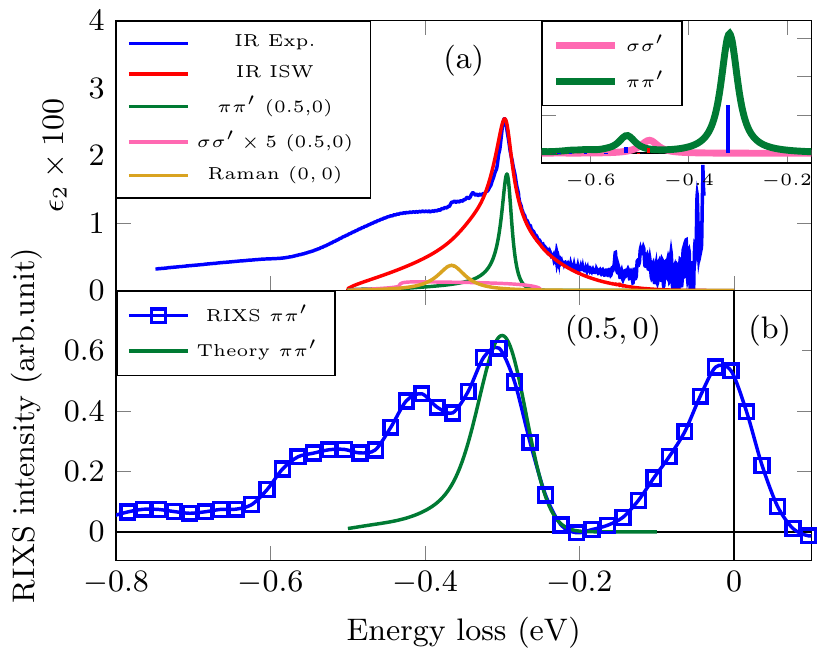}
\caption{(a) Blue shows the IR line shape from Ref.\,\onlinecite{Perkins1993} plotted as imaginary part of the dielectric function (assuming a dielectric constant $\epsilon_1=5$ and a phonon frequency shift of \SI{0.61}{\milli\electronvolt}). The other curves are the interacting spin-wave theory with  $J=\SI{0.108}{\electronvolt}$, see Refs.\,\onlinecite{Lorenzana1999,Lorenzana1995,LorenzanaSawatzky1995,DonkovChubukov2007} for details.
 (b) The  $\pi\pi'$ experimental RIXS lineshape and the theoretical two-magnon theory with the same energy position used to fit the IR spectra and an
experimental Gaussian broadening FWHM = \SI{65}{\milli\electronvolt}.
The inset shows the multimagnon response using exact diagonalization in the Heisenberg model in a 32 site cluster as
implemented in Ref.\,\onlinecite{Kawamura2017,Hoshi2021}. %This gives spectral weight in the region
 } \label{fig:lanczos}
\end{figure}

 Conversely,  the $\sigma\pi'$ polarization at $(\nicefrac{1}{2}, 0)$ does not fit this scenario. Indeed, one might expect  $\sigma\pi'$ and $\pi\sigma'$  (red lines) to be proportional to each other and, eventually, to
$S^{\perp}((\nicefrac{1}{2}, 0),\omega)$, which is clearly not the case. This implies that the scattering operator in RIXS, in addition to the standard on-site spin-flip process  $\hat{S}^\pm(\bf r)$ includes  non-local contributions that can be sensitive to the electric field (i.e., photon polarization) orientation. A possible explanation is a generalization of the three spin operator proposed in Ref.\,\cite{Ament2010},
$\hat{S}^\pm_{\bf r} \hat{\bf S}_{\bf r}\cdot\hat{\bf S}_{{\bf r}+\bm{\delta}}$ with a matrix element depending on the projection of the electric field on the bond direction $\bm{\delta}$. Interference between these two channels allows to rationalize the need for a Fano line shape for the fitting, and the different spectral shapes between RIXS and INS and between the $\pi\sigma'$ and $\sigma\pi'$ configurations. Multi-magnon scattering in RIXS was discussed before \cite{Igarashi2012} as part of the standard $S^{\perp}({\bf q},\omega)$. Here we  propose that the weight of these excitations can be modulated by the photon polarization (see SM\cite{SupplementaryMaterial}). 

We now turn to the $\Delta S=0$ excitations, which can be probed in RIXS as well as Raman and infrared (IR) spectroscopy. In all these cases the scattering operator involves two spin operators $B_{\bf r}^{\bm{\delta}}\equiv\hat{\bf S}_{\bf r}\cdot\hat{\bf S}_{{\bf r}+\bm{\delta}}$, and can access excitations with an even number of magnons. 
In RIXS the scattering operator is usually derived assuming that the main effect of the intermediate $2p^53d^{10}$ state
is to transiently eliminate one magnetic site. This local approximation \cite{Ament2010,Braicovich2009} yields a polarization independent lineshape of the even order multi-magnon spectrum. However, also in this case, we find that the lineshape at the X-point is strongly dependent on the polarization  (blue lines in Fig\,\ref{fig:polarimeter}), which again calls for non-local effects of the core hole. Here the polarization effects can be incorporated already at the leading  two-spin operator channel, which facilitates an explicit computation of the spectral shape. Versions of such polarization dependent operators have already been proposed for RIXS \cite{DonkovChubukov2007,Vernay2007}. We adopt the following form,  
%and closely related to the one appearing in %the theory of IR absorption and %Raman \cite{Lorenzana1995,LorenzanaSawatzky1995}
$A^{nl}({\bf q})=f({\bf q})\sum_{\mu} (\hat {\bf e}^i\cdot\bm{\delta})  (\bm{\delta}\cdot\hat {\bf e}^{o}) B^{\bm{\delta}}({\bf q})$. Here $\hat {\bf e}^{i,o}$ are the polarization
vectors of the incoming and outgoing photons and  $B^{\bm{\delta}}({\bf q})$ the Fourier transform of  $B_{\bf r}^{\bm{\delta}}$ with  $f({\bf q})$ a polarization independent
form factor. The resulting lineshape
is closely related to the theory of phonon-assisted multimagnon excitation, where the same associated spectral function appears but momentum integrated with a different form factor \cite{Lorenzana1995,LorenzanaSawatzky1995,Lorenzana1999}. 

Figure\,\ref{fig:lanczos}(a) shows the IR experiment \cite{Perkins1993,Perkins1998} together with an interacting spin-wave theory (ISWT) computation restricted to two magnons\cite{Lorenzana1995,LorenzanaSawatzky1995}. This explains the leading peak in terms of the momentum-integrated two-magnon response but misses substantial weight in higher order side-bands, which was thus assigned to four- and higher multi-magnon processes \cite{Perkins1993,Perkins1998,Lorenzana1995,LorenzanaSawatzky1995,Lorenzana1999}. We also show the two-magnon spectral function at specific reciprocal space points, corresponding to the RIXS (green and pink) or Raman (brown) lineshape. Upon momentum integration,  the IR lineshape is dominated by a two-magnon resonance dubbed \emph{the bimagnon}, which corresponds to the proposed RIXS lineshape in the $\pi\pi'$ channel at the X-point. 
Indeed, the theoretical bimagnon lineshape, whose energy is assigned by the IR experiment, explains fairly well the leading observed RIXS peak (panel b). It may appear surprising that the bimagnon has nearly the same energy as the single magnon (Fig.~\ref{fig:polarimeter}a). This is explained by the attractive magnon-magnon interaction and by the fact that the bimagnon has contributions from low-energy magnons whose individual momentum is away from the zone boundary. Strikingly, it is clear from Fig.~\ref{fig:lanczos} that both RIXS and IR leave a similar fraction of spectral weight in higher multi-magnon processes further supporting a common explanation. 

In Fig.\,\ref{fig:lanczos} (a) we show also the  $\sigma\sigma'$ RIXS two-magnon ISWT prediction, which gives a broad and very weak peak (pink, multiplied by 5 to make the curves visible) in  agreement with the absence of the $\approx 0.3$ eV bimagnon peak in the experimental $\sigma\sigma'$ spectrum shown in blue in Fig.~\ref{fig:polarimeter}(b). The structure at $\approx 0.5$ eV can be thus assigned to four- and higher multi-magnon process. Indeed, the inset in Fig.~\ref{fig:lanczos} shows that the dramatic difference between the two polarization combinations ($\sigma\sigma'$ and $\pi\pi'$) is qualitatively reproduced if instead of restricting to two-magnons we perform an exact computation in a small cluster. This treatment, however, underestimates the relative weight of high energy side bands. The same problem arises for the IR lineshape and was explained as due to finite size effects and a substantial four-ring exchange term in the Hamiltonian \cite{Lorenzana1999}, which was omitted here for simplicity.% Our parallel polarization measurements confirm the bimagnon resonance predicted in Ref.\,\cite{LorenzanaSawatzky1995} and show that multimagnon excitations play an important role; however a large fraction of the spectral weight remains unexplained both for RIXS and IR.

Taking full advantage of the additional information contained in the RIXS lineshape (both in parallel and crossed polarization channels) requires computations which include non-local effects in the scattering cross section and resonant effects in the matrix elements and selection rules, which we hope our work will stimulate. In the future, the present findings can be extended to other magnetic systems and doped cuprate compounds. In IR experiments a remarkably different situation was found for spin-1 2D antiferromagnetic square lattice, where four-magnon and higher order processes remain negligibly weak \cite{Lorenzana1995} and the two-magnon theory suffices to reproduce the experimental lineshape\cite{Kastner1998}. We can understand this drastic difference by noting that these computations are based on a $1/S$ expansion, which might pose convergence problems for $S<1$. Our results are further evidence that $S=\nicefrac{1}{2}$ systems belong to a different class and are characterized by proximity to more exotic ground states \cite{Anderson2004}, as also proposed earlier by analyzing optical\cite{Ho2001} and  INS studies \cite{Headings2010,Christensen2007,Tsyrulin2009}.

\begin{acknowledgments}
The experimental data were collected at the beam line ID32 of the European Synchrotron (ESRF) in Grenoble (France) using the ERIXS spectrometer designed jointly by the ESRF and the Politecnico di Milano. This work was supported by ERC-P-ReXS Project No. 2016-0790 of the Fondazione CARIPLO, Regione Lombardia and by MIUR Italian Ministry for Research through project PIK Polarix and PRIN Project No. 2017Z8TS5B. J.L. acknowledges financial support from Regione Lazio (L. R. 13/08) under project SIMAP.
The computation using the exact diagonalization method in this work has been done using the facilities of the Supercomputer Center, the Institute for Solid State Physics, the University of Tokyo.
We thank Riccardo Arpaia, Tom Devereaux, Kurt Kummer, Matteo Minola and Marco Moretti-Sala
for fruitful discussions. N.B.B. thanks Diamond Light Source, UK, for hosting him during the writing of this manuscript.
Computations were done with support through ISCRA Class C project 
HP10C72OM1.
\end{acknowledgments}

\bibliography{cuprates}
\end{document}